# Development of Large-area Lithium-drifted Silicon Detectors for the GAPS Experiment

M. Kozai, H. Fuke, M. Yamada, T. Erjavec, C. J. Hailey, C. Kato, N. Madden,
K. Munakata, K. Perez, F. Rogers, N. Saffold, Y. Shimizu, K. Tokuda and M. Xiao

*Abstract*— We have developed large-area lithium-drifted silicon (Si(Li)) detectors to meet the unique requirements of the General Antiparticle Spectrometer (GAPS) experiment. GAPS is an Antarctic balloon-borne mission scheduled for the first flight in late 2020. The GAPS experiment aims to survey low-energy cosmic-ray antinuclei, particularly antideuterons, which are recognized as essentially background-free signals from dark matter annihilation or decay. The GAPS Si(Li) detector design is a thickness of 2.5 mm, diameter of 10 cm and 8 readout strips. The energy resolution of <4 keV (FWHM) for 20 to 100 keV X-rays at temperature of -35 to -45 C, far above the liquid nitrogen temperatures frequently used to achieve fine energy resolution, is required. We developed a high-quality Si crystal and Li-evaporation, diffusion and drift methods to form a uniform Li-drifted layer. Guard ring structure and optimal etching of the surface are confirmed to suppress the leakage current, which is a main source of noise. We found a thin un-drifted layer retained on the *p*-side effectively suppresses the leakage current. By these developments, we succeeded in developing the GAPS Si(Li) detector. As the ultimate GAPS instrument will require >1000 10-cm diameter Si(Li) detectors to achieve high sensitivity to rare antideuteron events, high-yield production is also a key factor for the success of the GAPS mission.

## I. INTRODUCTION

GAPS is an Antarctic balloon-borne project that aims to observe low-energy (< 0.25 GeV/n) cosmic-ray antinuclei, particularly antideuterons, which are recognized as uniquely sensitive signatures of dark matter annihilation or decay in the Galactic halo [1-3]. GAPS consists of the silicon detector array and TOF (time-of-flight) counter which is made of plastic scintillator paddles and surrounding the silicon array. A low-energy antinucleus triggered by TOF and stopped in the silicon detector generates an excited exotic atom with the silicon atom. Then, characteristic X-rays, pions and protons are radiated in the decay and nuclear annihilation of the exotic atom. We can identify the antinucleus species and reject backgrounds by reconstructing the energies and tracks of these exotic atom products with the silicon detector array. It is also noted that the TOF system provides high-speed trigger and veto, as well as dE/dx and velocity measurement of incoming ionizing particle for additional particle identification.

The silicon detector array, thus, has an essential role in the GAPS detection scheme. It is required to slow and capture antinuclei, measure 20 to 100 keV X-rays with FWHM <4 keV energy resolution and detect tracks of incident particles and exotic atom products. Due to the limited power available to the cooling system on the balloon flight, the operation temperature is -35 to -45 C, significantly higher than that of typical silicon X-ray detectors. More than 10 m$^2$ of active Si area is required to achieve sensitivity to antideuteron fluxes as low as the order of 10$^{-6}$ [m$^{-2}$s$^{-1}$sr$^{-1}$(GeV/n)$^{-1}$] [4,5]. In order to provide sufficient depth to stop incoming antinuclei, high escape fractions for the characteristic X-rays, tracking efficiency for the incoming antinucleus and outgoing annihilation products, and low-power operation necessary for long-duration balloon flight, the active thickness, active area, and number of strips must be optimized.

To meet these requirements, we adopt Si(Li) detectors, which have a fabrication process that allows us to make a thick sensitive layer [6,7]. Our optimized flight detector design is a thickness of 2.5 mm, diameter of 10 cm and 8 strips. Over 90% of the thickness must be a sensitive layer and leakage current below 0.5 nA/cm$^2$ at temperature of -35 to -45 C is required to achieve the energy resolution of <4 keV (FWHM) necessary to distinguish X-rays from different incident antinuclei. High-yield, low-cost mass production of 1440 detectors are required to fill the large sensitive volume of the GAPS silicon detector array.

In the Si(Li) fabrication process, a thick intrinsic layer, which functions as a sensitive region under reverse bias voltage, is produced by compensating *p*-type Si with positive Li donor ions. Conventional fabrication processes of Si(Li) for X-rays, however, have not been proven with the large active area (>5 cm in diameter), relatively high operating temperature (well above liquid nitrogen temperature), and high production yield necessary for GAPS. Uniform Li drift in large-area Si(Li) fabrication has been difficult, and high leakage current, which is a main source of noise and degrades energy resolution, is also a key issue in the large-area and high-temperature Si(Li) development. Based on previous works on prototype detectors [8-11], we have developed a mass production process for large-area Si(Li) detectors, which are necessary to achieve high sensitivity to cosmic antinuclei in the first flight of GAPS, scheduled in late 2020.

M. Kozai and H. Fuke are with Institute of Space and Astronautical Science, Japan Aerospace Exploration Agency (ISAS/JAXA), Sagamihara, Kanagawa 252-5210, Japan (e-mail: kozai.masayoshi@jaxa.jp).
M. Yamada and K. Tokuda are with Sensor Device Business Unit, Device Department, Shimadzu Corporation, Atsugi, Kanagawa 243-0213, Japan.
T. Erjavec, K. Perez, F. Rogers and M. Xiao are with Massachusetts Institute of Technology, Cambridge, MA 02130, USA.
C. J. Hailey, N. Madden and N. Saffold is with Columbia University, New York, NY 10027, USA.
C. Kato and K. Munakata are with Department of Physics, Shinshu University, Matsumoto, Nagano 390-8621, Japan.
Y. Shimizu is with Kanagawa University, Yokohama, Kanagawa 221-8686, Japan.

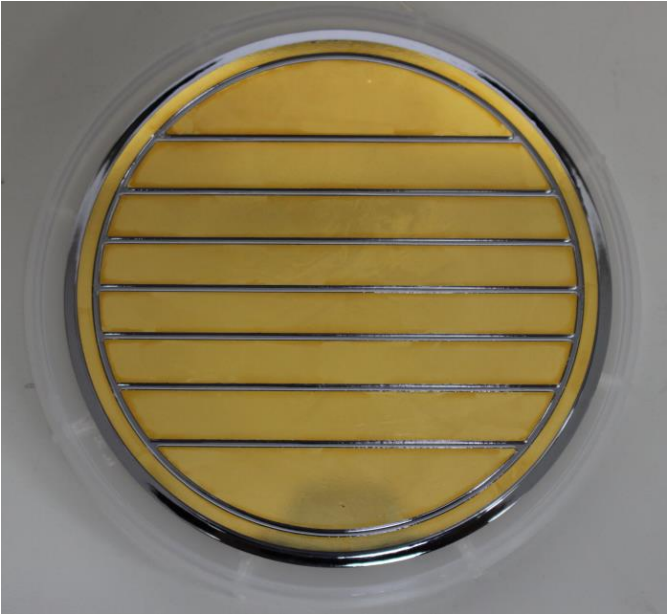

Figure 1 : Picture of a 2.5 mm-thick, 10 cm diameter Si(Li) detector with 8 readout strips. The $n+$ side on which Au is applied as a contact is divided into a guard ring (perimeter area) and 8 readout strips by grooves formed from ultrasonic impact grinding.

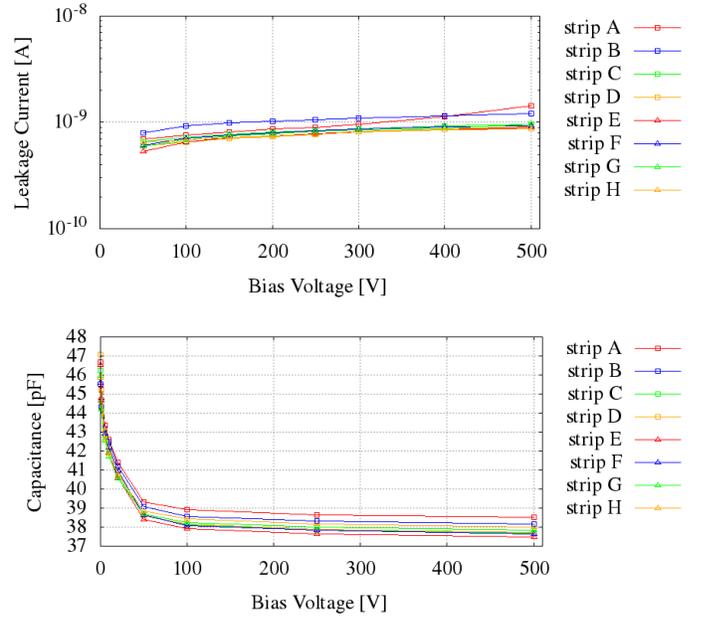

Figure 2 : Performance of all 8 strips of one sample Si(Li) detector at a temperature of -35 C. Upper panel shows Leakage current as a function of reverse bias voltage, while the lower panel shows capacitance as a function of reverse bias voltage.

## II. DEVELOPMENT OF THE DETECTOR

We describe below the development of a fabrication process for flight-geometry GAPS Si(Li) detectors. Figure 1 shows the $n$-side of an 8-strip detector produced via this process.

It previously has been difficult to make a uniform Li-drifted layer for thick and large-area Si(Li) detectors. This problem mainly arises from defects and contaminants such as oxygen or carbon in the Si crystal [12-15]. We have succeeded in obtaining a high-quality and $p$-type (Boron-doped) Si crystal from SUMCO Corporation, Japan and in producing a uniform Li-drifted layer with high yield, as showed below.

Li is first evaporated onto one face of a $p$-type Si wafer and thermally diffused through a short depth, forming an $n+$ layer of ~100 $\mu$m thickness. After removing the perimeter of the $n$-side by ultrasonic impact grinding (UIG), which forms a "top-hat" geometry that confines the Li drift region, Li ions are drifted through the bulk of the wafer by applying automatically controlled reverse bias and heater output. Li ions compensate B acceptor ions and other impurities in the $p$-type bulk, forming a thick intrinsic layer.

In large-area Si(Li) detectors, Joule heat generated by leakage current in the drift process is larger than in conventional, smaller Si(Li) detectors. We surveyed and deduced optimal parameters of bias voltage and heater controls which can successfully drift Li in a feasible period for mass production. It is also noted that our development of uniform heating methods in Li evaporation, diffusion and drift processes were key techniques to fabricate the uniform Li-drifted layer in addition to the high-quality $p$-type Si crystal.

In conventional Si(Li) fabrication techniques, the Li-drifted layer is exposed on the $p$-side, and a metal contact, such as Au, is used as a Schottky barrier to suppress leakage current. However, we observed excessively high leakage currents using this method, possibly indicating that a large-area Schottky barrier contact easily breaks down at high temperature. Based on this presumption, we optimized drift parameters so that a thin (~100 $\mu$m) un-drifted layer is retained on the $p$-side and confirmed that the leakage current is sufficiently suppressed. The uniform Li-drift and successful retention of an un-drifted layer were visually confirmed by copper-staining [16-18] on cross sections of Si(Li) detectors. We note that a Schottky barrier is typically adopted in order to minimize the $p$-side dead layer in detectors where the $p$-contact is a window for low-energy X-rays [19-21]. However, since GAPS aims to detect X-rays with energies higher than 20 keV, the ~100 $\mu$m dead layer is acceptable in the GAPS detection scheme.

After Li drifting, a circular groove of ~300 $\mu$m depth is cut into the $n$-side by UIG, so that the central area, or the readout area of the detector is electrically isolated from the perimeter area, or the guard ring [22]. Leakage current flowing into the readout circuit can be suppressed by this isolation, because the side surface of the top-hat is a main source of surface leakage current. We have verified that a guard ring structure drastically reduces the leakage current by a factor of at least $10^{-2}$ to $10^{-3}$. Grooves that electrically isolate the readout strips are cut at the same time as machining of the guard ring groove.

Chemical etching with a solution of hydrofluoric acid, nitric acid and acetic acid is performed between each fabrication stage to smooth the surface, remove impurities, and set the proper surface state. We found the last etching of the guard ring and strip grooves is essential to reduce the surface leakage current in the final detector. Too many etchings, on the other hand, result in expansion of the grooves, which not only decreases active area but also possibly increases leakage currents generated in groove surfaces. We therefore optimized the etching time that sufficiently reduces the reducing leakage current, by measuring leakage currents as a function of total etching time.

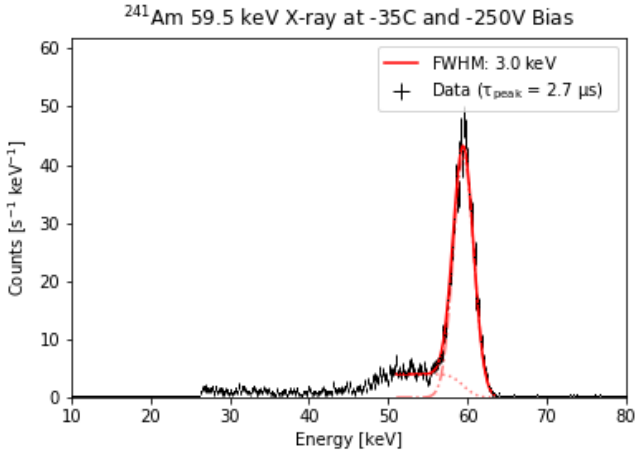

Figure 3 : Energy spectrum at a temperature of -35 C measured with one strip of a sample GAPS Si(Li) detector and 59.5 keV X-rays from [241]Am.

## III. Performance of the GAPS Si(Li) detectors

Figure 2 shows the leakage current and capacitance as a function of reverse bias voltage for one sample 8-strip detector at an operating temperature of -35 C. The leakage current of each strip is ~1 nA at our operating voltage (200 to 300 V). This corresponds to ~0.1 nA/cm$^2$, significantly lower than 0.5 nA/cm$^2$ that is required to achieve 4 keV energy resolution for 20 to 100 keV X-rays.

The capacitance reaches a minimum by ~100 V, indicating the detector is fully depleted around this voltage. Indeed, the capacitance of $C \sim 38$ pF corresponds to a thickness of depletion layer [7],

$$W \sim 1.05 A/C \sim 2.4 \text{ mm}$$

with a strip area of $A = 8.7$ cm$^2$. This is 96% of the detector thickness, 2.5 mm. The uniformity of the capacitance, with a fluctuation less than ~3 % between strips, indicates a uniform Li-drifted layer.

Figure 3 displays the energy spectrum at a temperature of -35 C measured with one strip of a sample Si(Li) detector and 59.5 keV X-rays from [241]Am. This is a preliminary result, but we have confirmed the required X-ray energy resolution, <4 keV (FWHM), can be achieved by our detectors.

Evaluation of the yields for this fabrication process are ongoing, but >90% of strips of the most recent ~10 detectors demonstrate leakage current and capacitance that satisfy our requirements.

## IV. Conclusion

We have developed a high-yield fabrication method for large-area Si(Li) detectors operated at relatively high temperature for the GAPS experiment. We developed Li evaporation, diffusion and drift apparatuses and successfully optimized the key parameters of bias voltage and heater controls in the Li-drift. We also confirmed that a thin un-drifted layer and the guard ring structure are highly effective to reduce the leakage current for these large-area Si(Li) detectors.

The yield rate of detectors meeting the GAPS requirements is sufficiently high at this point. Their leakage currents are low enough to achieve <4 keV energy resolution for 20 to100 keV X-rays, and their capacitances are sufficiently low, indicating the detectors are fully depleted. The leakage current and capacitance as a function of bias is similar between all strips, indicating that the detector volume is uniformly compensated by our Li-drift. This is also confirmed by copper-staining on cross sections of sample Si(Li) detectors.

We measured the energy spectrum of a sample 8-strip detector and confirmed energy resolution below 4 keV for 59.5 keV X-rays at temperature of -35 C.

We will start mass production of GAPS Si(Li) detectors in late 2018, extending through early 2020.

## Acknowledgement

We thank SUMCO Corporation and Shimadzu Corporation for their co-operations in detector development.

M. Kozai is supported by the JSPS KAKENHI under Grant No. JP17K14313. H. Fuke is supported by the JSPS KAKENHI under Grant No. JP2670715 and JP17H01136. K. Perez receives support from the Heising-Simons Foundation and the Alfred P. Sloan Foundation. F. Rogers is supported through the National Science Foundation Graduate Research Fellowship under Grant No. 1122374.

This work was partially supported by the NASA APRA program through grant NNX17AB44G.